\documentclass[aps,twocolumn,showpacs,superscriptaddress,amsmath,amssymb,amsfonts,floatfix,longbibliography]{revtex4-2}

\usepackage[T1]{fontenc} 
\usepackage{graphicx}
\usepackage{float}
\usepackage{dcolumn}
\usepackage{bm}
\usepackage{amssymb}
\usepackage{tabularx}
\usepackage{microtype}
\usepackage{xfrac}
\usepackage{chemformula}
\usepackage{gensymb}
\usepackage[most]{tcolorbox}
\usepackage{xcolor}
\usepackage{multirow}
\usepackage{enumitem}
\usepackage{array}
\usepackage{booktabs}
\usepackage[flushleft]{threeparttable}
\usepackage{natbib,hyperref}
\usepackage{graphicx}
\usepackage{makecell}
\usepackage{colortbl}
\usepackage{xcolor}
\usepackage{soul}

\hypersetup{
	colorlinks=true, %set true if you want colored links
	citecolor=blue,  %choose some color if you want links to stand out
}

\begin{document}

\title{More is different: Case studies of how chemical complexity\\influences stability in high entropy oxides}
%Beyond entropy: Chemical factors influencing the stability of high entropy oxides}

\author{Ksenia Khoroshun}
    \affiliation{Stewart Blusson Quantum Matter Institute, University of British Columbia, Vancouver, BC V6T 1Z4, Canada}
    \affiliation{Department of Physics \& Astronomy, University of British Columbia, Vancouver, BC V6T 1Z1, Canada}

\author{Mario U. Gonz\'alez-Rivas}
    \affiliation{Stewart Blusson Quantum Matter Institute, University of British Columbia, Vancouver, BC V6T 1Z4, Canada}
    \affiliation{Department of Physics \& Astronomy, University of British Columbia, Vancouver, BC V6T 1Z1, Canada}

\author{Alannah M. Hallas}
\email[Email: ]{alannah.hallas@ubc.ca}
    \affiliation{Stewart Blusson Quantum Matter Institute, University of British Columbia, Vancouver, BC V6T 1Z4, Canada}
    \affiliation{Department of Physics \& Astronomy, University of British Columbia, Vancouver, BC V6T 1Z1, Canada}
    \affiliation{Canadian Institute for Advanced Research (CIFAR), Toronto, ON, M5G 1M1, Canada}

\date{\today}
\begin{abstract}
\smallskip
\begin{center}
{\normalsize\textbf{ABSTRACT}}\\    
\end{center}

Tailoring the chemical composition of a high entropy oxide (HEO) is a powerful approach to enhancing desirable material properties. However, the targeted synthesis of HEO materials is often hindered by competing stabilizing and destabilizing factors, which are difficult to predict. This work examines the effects of increased configurational entropy on the phase formation and stability of four notable complex oxide families: perovskite ($AB$O\textsubscript{3}), pyrochlore ($A$\textsubscript{2}$B$\textsubscript{2}O\textsubscript{7}), Ruddlesden-Popper ($A_{2}B$O\textsubscript{4}), and zirconium tungstate ($AB$\textsubscript{2}O\textsubscript{8}). Each of these structures has a tetravalent cation site, which we attempt to substitute with an entropic mixture of four cations, benchmarked by the parallel synthesis of a \textcolor{black}{non-disordered reference compound}. While all four target high entropy materials can be expected to form based on ionic radii criteria, only the high entropy perovskite Ba(Ti,Zr,Hf,Sn)O$_3$ is successfully synthesized. In the case of the pyrochlore, an entropy-stabilized defect fluorite is formed instead, while the Ruddlesden-Popper phase co-exists with multiple competing phases. For the tungstate, an unexpected deep eutectic point between the precursors results in melting that precedes the formation of a high entropy phase. 
Our case studies therefore illustrate that the stability of HEOs cannot be straightforwardly predicted based on ionic radii, lattice geometry, and charge-balancing considerations alone due to the underlying complexity of the interactions between the many chemical constituents. 

\end{abstract}

\maketitle

\section*{\label{sec:Introduction}Introduction}

In recent years, high entropy oxides (HEOs) have emerged as a promising materials platform for tailored physical and chemical properties \cite{Oses2020, Musicó2020, mccormack2021thermodynamics, Solveig, kotsonis2023high, Sen2024, Schweidler2024}. The first HEO, a rock salt structured material, was reported by Rost \emph{et al.} in 2015~\cite{Rost2015}. Following this, high entropy versions of various oxide crystal structures have been synthesized and studied \cite{Sen2024, Musicó2020}. Remarkably, increasing the configurational entropy of an oxide has been shown to enhance certain physical properties relative to their low entropy counterparts, such as improved battery cycling stability \cite{HEO_battery,lun2021cation}, increased dielectric breakdown strength \cite{Jing2023, Fan2025}, and more effective performance as a thermal barrier material \cite{li2019high,ZHAO202045}. Therefore, tailoring the chemical composition of existing oxide structures by increasing their configurational entropy appears to be a promising route to enhance material properties or to discover new emergent properties.

There is no universal definition for what constitutes an HEO~\cite{brahlek2022name,Solveig,fracchia2024phase}. Some works favor the computed value of the nominal Boltzmann entropy, which presupposes an ideally disordered cation arrangement free from short-range order or clustering, while other works emphasize the role of entropy stabilization. Here, we adopt a holistic definition that a material has meaningfully ``high entropy'' when the configurational entropy significantly exceeds that of a conventional solid solution and when the cation disorder is observed to modify the formation or physical properties of the resulting material. Accordingly, a material with fewer than five cations can still be considered high entropy if, for example, the configurational entropy is responsible for selecting the crystal structure or leads to a desirable functional property.

Given the vast number of complex oxide structures available to be studied and the combinatorial complexity of cations that may be incorporated into a single lattice site, understanding the stability trends of HEOs represents a monumental challenge. While the stability of many conventional oxide structures can be rationalized in terms of local electrostatics, ionic radius, and lattice geometry~\cite{pauling1929principles}, these heuristics can break down in the case of HEOs. One interesting example comes from the $R_2$TiO$_5$ ($R=$~rare earth) family of HEOs where the boundaries between different structure types are observed to significantly vary from what is expected based on average ionic radius alone~\cite{HE_RE_complex_oxides,oquinn2021defining}. %Meanwhile, work on high entropy rare earth iridate pyrochlores,  $R_2$Ir$_2$O$_7$, demonstrated that the electronic phase transition from metal to Mott insulator correlates precisely with the average ionic radius~\cite{contant2024robust}. 
Studies on perovskites, $AB$O$_3$ have shown a high success rate in the synthesizability of compositions with $B$-site cation valences ranging from 2+ to 6+ with large deviations in ionic radius, so long as the average is strictly held to 4+~\cite{ma2021high,tang2021high}. This contrasts with the case of simple tetravalent HEOs $A$O$_2$, where only a single high entropy mixture out of 56 was found to be synthesizable~\cite{Aamlid2023}. Work on rock salt HEOs, $A$O, has shown that in cases of competing valence states, precise control of oxygen partial pressure is needed to stabilize a single-phase material~\cite{almishal2025thermodynamics}. The formation of high entropy phases is therefore restricted by the interplay of competing stabilizing and destabilizing effects, which are often difficult to predict.   

This work aims to explore the impact of increased entropy on phase formation and stability in four families of complex oxides. Using the same mixture of tetravalent cations (Ti, Zr, Hf, and Sn) and working within the anticipated structural tolerance window for each structure, we directly compare the synthesis results of four high-entropy compounds with their \textcolor{black}{non-disordered reference compounds}. We find that while some structures straightforwardly accept the entropic mixture, others will form a modified structure or fail to form altogether. In one case, an unexpected eutectic between the precursors results in a melting transition that precludes phase formation. From these observations, we conclude that one cannot straightforwardly predict the formation of HEOs through heuristics alone and that a more holistic approach is necessary to accurately describe the stability trends of these materials.

\section*{Results and Discussion}

\subsection*{Materials Theory}

The complex oxide families examined here belong to the perovskite ($A^{2+}B^{4+}$O$_3$), pyrochlore ($A^{3+}_2B^{4+}_2$O$_7$), Ruddlesden-Popper ($A^{2+}_2B^{4+}$O$_4$), and zirconium tungstate ($A^{4+}B^{6+}_2$O$_8$) structures. These materials families were chosen to fulfill several criteria: (i) with the exception of zirconium tungstate, they are common structure types that occur for many different combinations of cations; (ii) they all possess a nominally tetravalent cation site, which is the site we will attempt to substitute with an entropic mixture; and (iii) that tetravalent cation has octahedral oxygen coordination, with varying levels of distortion.

The cation mixture that we attempt to incorporate into each of these structures -- Ti, Zr, Hf, and Sn -- all exist in a stable 4+ oxidation state. Previous work has established that these four cations can co-exist in a binary oxide structure. Furthermore, (Ti,Zr,Hf,Sn)O\textsubscript{2} is entropy stabilized and has negligible short-range cation ordering~\cite{he2021four,Aamlid2023,entropy_mix}. Therefore, this cation combination is a good candidate to attempt to incorporate into more complex oxide structures. The Shannon radii of these four cations in a six-fold coordination environment are detailed in Table \ref{tab:radii}, where we see that the average ionic radius of the four is $r_{\text{avg}}=0.68$~\AA. %In the case of the Ruddlesden-Popper and the zirconium tungstate structures, we have also attempted to incorporate Ru, which has an ionic radius of 0.62~\AA, well within the range of the other four cations. 
For each structure type, we also select a known  \textcolor{black}{non-disordered reference phase} to benchmark our synthesis protocol and structural characterization. Where possible, we use Sn as its ionic radius is closest to the average value. \textcolor{black}{Throughout the manuscript, green checks and red crosses are used in the figures to mark successful and unsuccessful synthesis outcomes, respectively.}

\begin{table}[h]
    \centering
      \renewcommand{\arraystretch}{1.2}
    \begin{tabular}{|c|c|c|}
    \hline
       Cation  & Z & Six-fold coordination radius \\
       \hline 
        Ti\textsuperscript{4+} & 22 & 0.605 \AA \\
        Zr\textsuperscript{4+}& 40 & 0.72 \AA \\
        %Ru\textsuperscript{4+} & 44 & 0.62 \AA \\
        Sn\textsuperscript{4+} & 50 & 0.69 \AA \\
        Hf\textsuperscript{4+} & 72 & 0.71 \AA \\
   \hline \hline
        Cation mix & Average Z & Average radius \\
        \hline
        Ti, Zr, Hf, Sn & 46 & 0.68(5) \AA\\ 
%        Ti, Zr, Hf, Sn, Ru & 45.6 & 0.67(5) \AA\\ 
 %       Zr, Hf, Sn, Ru & 51.5 & 0.69(4) \AA\\
 \hline          
         
    \end{tabular}
    \caption{Shannon radii of the four cations incorporated in the high entropy materials. Only 6-fold coordination is considered, as all tetravalent cation sites in the chosen oxide families exist in a 6-fold coordination. Note that the radius of tin is closest to the average ionic radius of the four cations. Values obtained from \cite{shannon}.}
    \label{tab:radii}
\end{table}

\subsection*{Perovskite Structure}

\begin{figure*}
\centering
\includegraphics[width=\textwidth]{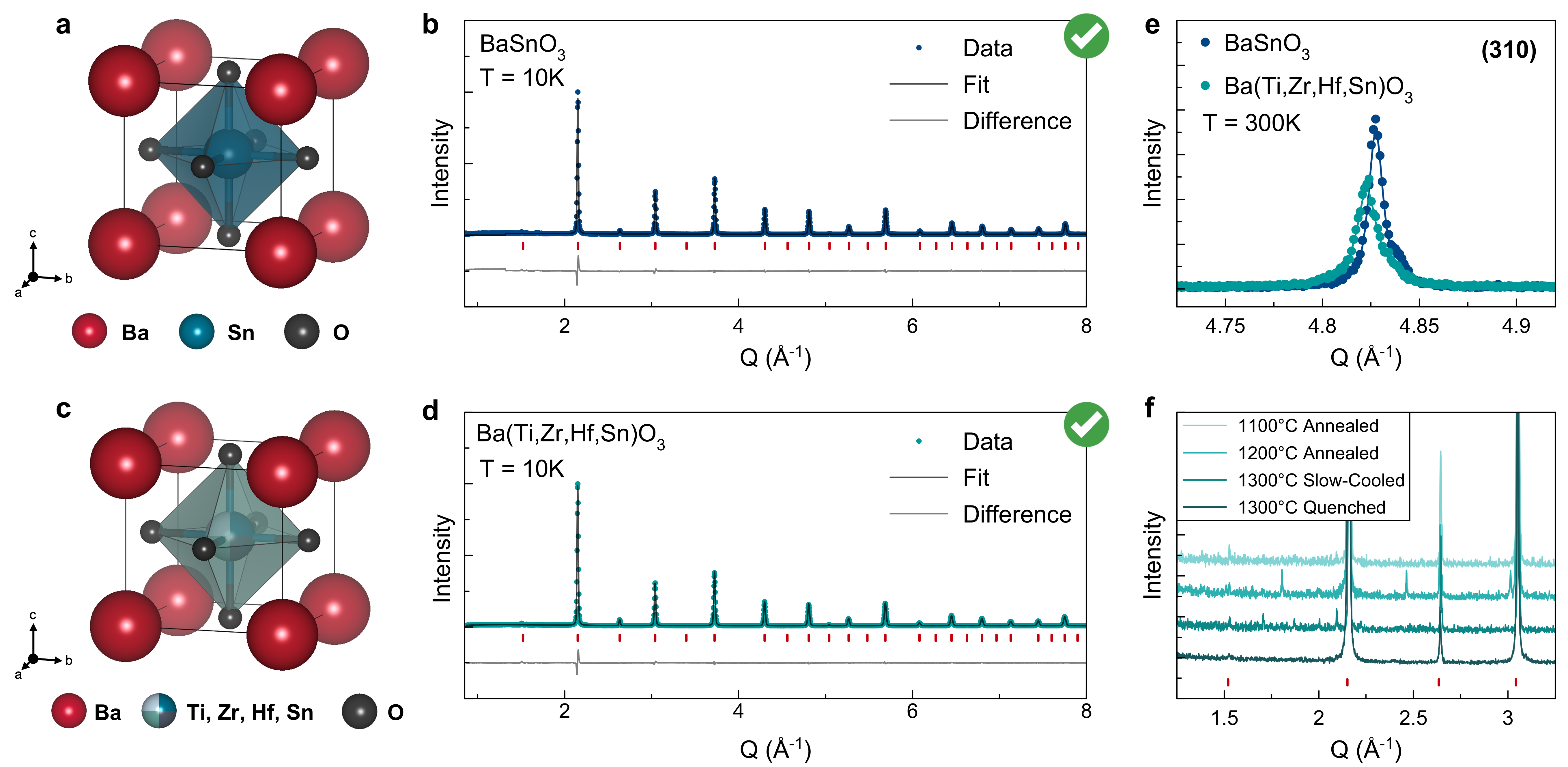}
 \caption{\textbf{Successful synthesis of a high entropy perovskite.} \textbf{(a)} The \textcolor{black}{reference} cubic perovskite BaSnO$_3$ with space group $Pm\overline{3}m$ was successfully prepared as a single-phase material via solid state synthesis, as shown by \textbf{(b)} Rietveld refinement of the \textcolor{black}{synchrotron} powder XRD pattern (\textcolor{black}{$R_{\text{wp}} = 9.10\%$}). \textbf{(c)} The high entropy perovskite, Ba(Ti,Zr,Hf,Sn)O\textsubscript{3} was also successfully synthesized as confirmed by the \textbf{(d)} Rietveld refinement of its \textcolor{black}{synchrotron} powder XRD pattern in the same $Pm\overline{3}m$ structure (\textcolor{black}{$R_{\text{wp}} = 4.61\%$}). \textbf{(e)} Comparison of the peak shape for the (310) Bragg peak in BaSnO$_3$ and Ba(Ti,Zr,Hf,Sn)O\textsubscript{3} \textcolor{black}{for data collected on a laboratory diffractometer with a Cu $K_{\alpha}$ source}. The two patterns have been normalized according to the highest intensity (110) Bragg peak. Disorder effects and strain result in a suppression of the total intensity and peak broadening in the case of the high entropy analog. \textbf{(f)} Thermal stability of Ba(Ti,Zr,Hf,Sn)O\textsubscript{3} showing the formation of weak impurity peaks when slow cooled or annealed below the synthesis temperature. Minimal impurities are observed at 1100\textdegree C, indicating kinetic stability at that temperature.}
\label{fig:Perovskite_doublecol}
\end{figure*}

The first class of materials we consider is the perovskites, $AB$O$_3$, which are among the most common and structurally well-understood families of complex oxides. Their phase stabilities and distortions are accurately predicted by simple tolerance factors, based on the ionic radii of the constituent elements~\cite{goldschmidt1926gesetze,bartel2019new}. We selected BaSnO$_3$ as our  \textcolor{black}{reference phase} and Ba(Ti,Zr,Hf,Sn)O\textsubscript{3} as our high entropy variant, which is well within the expected perovskite stability range based on both the average and individual ionic radii of the $B$-site cations~\cite{bartel2019new}. Employing conventional solid-state synthesis, both materials were successfully synthesized and found to crystallize in the cubic perovskite structure, as shown in Fig.~\ref{fig:Perovskite_doublecol}(a,c). 

To understand the effect of high entropy on the perovskite structure, we compare the structural parameters of the high entropy and \textcolor{black}{non-disordered reference} $AB$O$_3$ samples obtained via Rietveld refinement of their \textcolor{black}{synchrotron} powder x-ray diffraction (XRD) patterns, collected at $T=10$~K, the results of which are shown in Fig.~\ref{fig:Perovskite_doublecol}(b,d) and the refined parameters are presented in Table~\ref{tab:BaHEO3_refinement}. Excellent agreement was found for both BaSnO\textsubscript{3} and Ba(Ti,Zr,Hf,Sn)O\textsubscript{3} \textcolor{black}{($R_{\text{wp}} = 9.10\%$ and $R_{\text{wp}} = 4.61\%$, respectively)} modeled within space group $Pm\overline{3}m$ (no. 221). The cubic lattice parameters of the two unit cells are very similar (\textcolor{black}{$a = 4.12954(7)$~\AA\ }in the case of BaSnO\textsubscript{3} and $4.13374(6)$~\AA\ in the case of Ba(Ti,Zr,Hf,Sn)O\textsubscript{3}), as expected given that the average ionic radius of the high entropy cation mix (0.68~\AA) is comparable to the ionic radius of Sn alone (0.69 \AA). 

The most significant structural difference between the two materials is the broadening and reduced intensity of the higher angle Bragg peaks in Ba(Ti,Zr,Hf,Sn)O\textsubscript{3} relative to BaSnO\textsubscript{3}. \textcolor{black}{To quantify this effect, we refined the microstrain broadening for both samples, which is a measure of the root mean square (RMS) spread in interplanar distance and hence, a measure of the lattice parameter spread across each system. This analysis for BaSnO\textsubscript{3} revealed a microstrain corresponding to a 0.06\% RMS interplanar distance modulation. In contrast, Ba(Ti,Zr,Hf,Sn)O\textsubscript{3} exhibits a 0.2\% modulation in RMS interplanar distance, an order of magnitude difference, highlighting the impact of the high entropy substitution on the host perovskite structure.}

\textcolor{black}{Further effects directly ascribable to the high entropy substitution are observed in the thermal displacement parameters extracted from the Rietveld refinement of the reference and high entropy perovskite oxides. The Ba and O sublattices of BaSnO\textsubscript{3} and Ba(Ti,Zr,Hf,Sn)O\textsubscript{3} present differing identical thermal displacement behavior, disagreeing only in the last decimal place. Yet, the $U_{iso}$ value for the $B$ sublattice in Ba(Ti,Zr,Hf,Sn)O\textsubscript{3} is nearly twice that of Ba(Ti,Zr,Hf,Sn)O\textsubscript{3}, likely reflecting the incoherent atomic vibrations that the high entropy substitution induces into the perovskite structure.}

\renewcommand{\belowrulesep}{0pt}
\renewcommand{\aboverulesep}{0pt}
{\setlength{\tabcolsep}{0pt}
\begin{table}[tbp]
\caption{Refined structural parameters for perovskites \textcolor{black}{BaSnO$_3$ and} Ba(Ti,Zr,Hf,Sn)O\textsubscript{3} in the $Pm\overline{3}m$ space group \textcolor{black}{from synchrotron diffraction data collected at ID-28-1 at $T=10$~K.}  }
\centering
%\resizebox{\columnwidth}{!}{%
\renewcommand{\arraystretch}{1.1} % Increases row height by 50%
\begin{tabular}{lcccccc}
\toprule
\rowcolor[HTML]{DADADA} 
\multicolumn{1}{c}{\cellcolor[HTML]{DADADA}} &
  \cellcolor[HTML]{DADADA} &
  \multicolumn{3}{c}{\cellcolor[HTML]{DADADA}\textbf{Coordinates}} &
  \cellcolor[HTML]{DADADA} &
  \cellcolor[HTML]{DADADA} \\
\rowcolor[HTML]{DADADA} 
\multicolumn{1}{c}{\multirow{-2}{*}{\cellcolor[HTML]{DADADA}\textbf{~~Atom~~}}} &
  \multirow{-2}{*}{\cellcolor[HTML]{DADADA}\textbf{~~Site~~}} &
  \textbf{~~x~~} &
  \textbf{~~y~~} &
  \textbf{~~z~~} &
  \multirow{-2}{*}{\cellcolor[HTML]{DADADA}\textbf{~~Occ.~~}} &
  \multirow{-2}{*}{\cellcolor[HTML]{DADADA}\textbf{\(~\mathbf{U_{iso}}~\)(\AA \(\mathbf{^2}\)) \hspace{1pt} } } \\ 
  \hline
  \multicolumn{7}{c}{BaSnO\textsubscript{3}~~~$a = 4.12954(7)$~\AA}\\
  \hline

Ba1 & $1b$                   & 0.5                    & 0.5                    & 0.5                    & 1        & 0.0052(4)  \\
\rowcolor[HTML]{DADADA} 
Sn1 &    $1a$                  &      0                       &             0                &          0                   &                 1            &     0.0019(4)    \\
O1 & $3d$                  & 0.5                    & 0                    & 0                    & 1        & 0.009(1) \\
\toprule
\multicolumn{7}{c}{~~}\\
\toprule
\rowcolor[HTML]{DADADA} 
\multicolumn{1}{c}{\cellcolor[HTML]{DADADA}} &
  \cellcolor[HTML]{DADADA} &
  \multicolumn{3}{c}{\cellcolor[HTML]{DADADA}\textbf{Coordinates}} &
  \cellcolor[HTML]{DADADA} &
  \cellcolor[HTML]{DADADA} \\
\rowcolor[HTML]{DADADA} 
\multicolumn{1}{c}{\multirow{-2}{*}{\cellcolor[HTML]{DADADA}\textbf{~~Atom~~}}} &
  \multirow{-2}{*}{\cellcolor[HTML]{DADADA}\textbf{~~Site~~}} &
  \textbf{~~x~~} &
  \textbf{~~y~~} &
  \textbf{~~z~~} &
  \multirow{-2}{*}{\cellcolor[HTML]{DADADA}\textbf{~~Occ.~~}} &
  \multirow{-2}{*}{\cellcolor[HTML]{DADADA}\textbf{\(~\mathbf{U_{iso}}~\)(\AA \(\mathbf{^2}\)) \hspace{1pt} } } \\ 
  \hline
\multicolumn{7}{c}{Ba(Ti,Zr,Hf,Sn)O\textsubscript{3}~~~  $a = 4.13443(4) $~\AA}\\
\hline
Ba1 & $1b$                   & 0.5                    & 0.5                    & 0.5                    & 1        & 0.0059(3)  \\
\rowcolor[HTML]{DADADA} 
Ti1 &                      &                             &                             &                             &                             &             \\
\rowcolor[HTML]{DADADA} 
Zr1 &                      &                             &                             &                             &                             &             \\
\rowcolor[HTML]{DADADA} 
Hf1 &                      &                             &                             &                             &                             &             \\
\rowcolor[HTML]{DADADA} 
Sn1 & \multirow{-4}{*}{$1a$} & \multirow{-4}{*}{0}     & \multirow{-4}{*}{0}     & \multirow{-4}{*}{0}     & \multirow{-4}{*}{0.25}   & \multirow{-4}{*}{0.0032(3)}     \\

O1 & $3d$                  & 0.5                    & 0                    & 0                    & 1        & 0.0089(8) \\ \bottomrule
\end{tabular}%

\label{tab:BaHEO3_refinement}
\end{table}
}

In order to assess the thermal stability of Ba(Ti,Zr,Hf,Sn)O\textsubscript{3}, a portion of the sample was re-annealed at its formation temperature (1300\textdegree C) and slowly cooled at a rate of 10\textdegree C an hour, as opposed to quenching. This slow-cooling largely preserved the cubic perovskite structure, but did lead to the formation of some weak impurity peaks, as shown in Fig.~\ref{fig:Perovskite_doublecol}(f). More prominent impurities were observed upon re-annealing the sample below the synthesis temperature at 1200\textdegree C for 64 hrs, while the perovskite Bragg reflections were notably broadened. The same treatment at still lower temperature, 1100\textdegree C for 100 hrs, did not result in significant degradation. Taken altogether, we can conclude that Ba(Ti,Zr,Hf,Sn)O\textsubscript{3} may be weakly entropy stabilized and is kinetically stable at temperatures of 1100\textdegree C and lower.

Despite the disorder-induced strain, the perovskite structure is clearly able to withstand a high level of configurational entropy, as has been found for a host of previously synthesized materials. These include $AB$O$_3$-type HEOs with entropic mixtures on the $A$-site~\cite{witte2020magnetic,zhang2023formation}, the $B$-site~\cite{jiang2018new,sharma2018single,tang2021high,ma2021high}, and mixtures across both cation sublattices~\cite{sarkar2018rare,witte2019high}. It is interesting to note that a previous study that considered the identical composition studied here found that Ba(Ti,Zr,Hf,Sn)O$_3$ could not be stabilized as a single-phase material~\cite{jiang2018new}. The reason for the discrepancy is unclear as the reported synthesis conditions are similar to those used here; while we obtained a single-phase material after sintering at 1200\textdegree C and then 1300\textdegree C, the previous study started at 1300\textdegree C and went to a maximum of 1500\textdegree C where impurity levels were observed to increase. It is therefore possible that Ba(Ti,Zr,Hf,Sn)O$_3$ is thermodynamically unstable at higher temperatures, again suggestive that entropy stabilization is likely not a significant factor. The same work identified that a single-phase perovskite could be formed with the addition of a fifth element (either Ce, Y, or Nb)~\cite{jiang2018new}. Taken together, these results indicate that, beyond geometric and electrostatic constraints, carefully optimized synthesis conditions are critical to stabilizing a targeted HEO phase.

\begin{figure*}[htb]
\centering
\includegraphics[width=\textwidth]{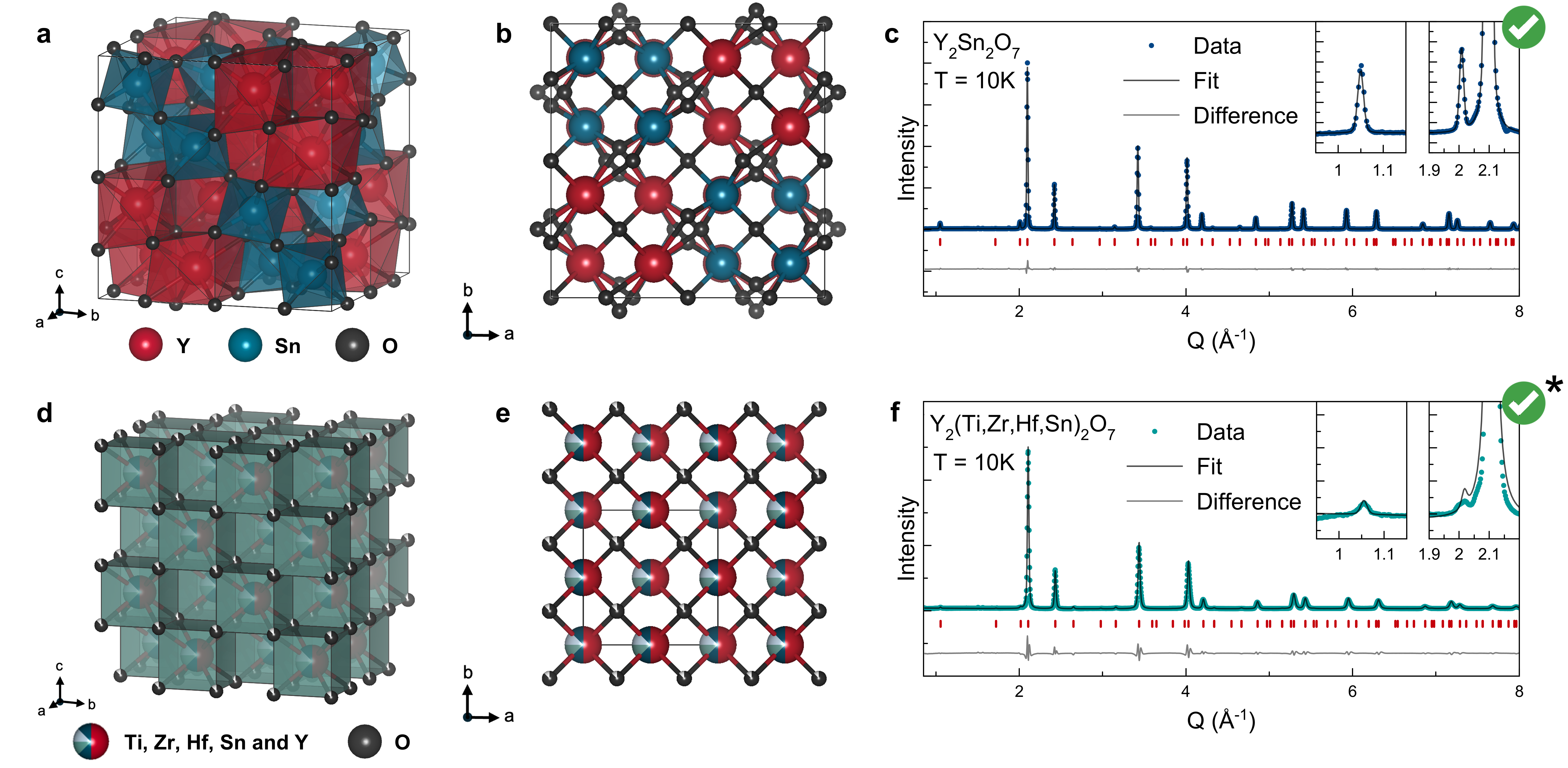}
 \caption{\textbf{Formation of an entropy-selected defect fluorite pyrochlore phase.} \textbf{(a,b)} Crystal structure of the \textcolor{black}{reference pyrochlore compound}  Y$_2$Sn$_2$O$_7$, where Y and Sn occupy distorted cubic and octahedral oxygen environments, respectively. \textbf{(c)} Rietveld refinement of the \textcolor{black}{synchrotron} powder XRD pattern for Y$_2$Sn$_2$O$_7$ confirming the formation of a pyrochlore phase in the cubic $Fd\overline{3}m$ space group (\textcolor{black}{$R_{\rm{wp}} = 7.34\%$}). The inset shows the (111) and (311) superlattice peaks at 1 and 2~\AA$^{-1}$, respectively. \textbf{(d,e)} Crystal structure of the defect fluorite-structured high entropy phase Y$_2$(Ti,Zr,Hf,Sn)$_2$O$_7$, where all the cations share a single crystallographic site with cubic coordination to a partially vacant oxygen sublattice. \textbf{(f)} Rietveld refinement of the \textcolor{black}{synchrotron} powder XRD pattern for Y$_2$(Ti,Zr,Hf,Sn)$_2$O$_7$ in the cubic $Fd\overline{3}m$ space group (\textcolor{black}{$R_{\rm{wp}}=12.01\%$}). The superlattice reflections, as shown in the inset, are almost fully suppressed, indicating substantial site mixing that more closely resembles a defect fluorite structure.}
\label{fig:pyrochlore}
\end{figure*}

\subsection*{Pyrochlore and Fluorite Structures}

The second family of complex oxides we consider is pyrochlores, $A_2B_2$O$_7$, a structure with an extensive stability range for many combinations of cations~\cite{subramanian1983oxide}. The defining structural characteristic of pyrochlore materials is the site-ordered $A$ and $B$ sublattices, both of which independently form a network of corner-sharing tetrahedra. \textcolor{black}{The $A$-site cations sit in an 8-fold coordinate distorted cubic oxygen environment while the $B$-site cations occupy a distorted octahedral environment} as shown in Fig.~\ref{fig:pyrochlore}(a,b) for the \textcolor{black}{reference} pyrochlore Y$_2$Sn$_2$O$_7$. This material was successfully obtained via solid state synthesis as shown by the Rietveld refinement of the powder XRD pattern in the $Fd\overline{3}m$ (no. 227) space group (Fig.~\ref{fig:pyrochlore}(c)). The characteristic pyrochlore superlattice peaks are visible at approximately 1 and 2~\AA$^{-1}$, as emphasized in the inset. 

The targeted high-entropy pyrochlore phase, Y\textsubscript{2}(Ti,Zr,Hf,Sn)\textsubscript{2}O\textsubscript{7}, was also successfully synthesized as a single-phase material, and its synchrotron powder diffraction pattern is presented in Fig.~\ref{fig:pyrochlore}(f). However, the intensities of the superlattice peaks for the high entropy phase are dramatically diminished: they can only be resolved in datasets with extremely high signal-to-noise ratios, such as those achievable at synchrotron light sources, and were not resolvable on our lab-based diffractometer. These extremely weak superlattice peaks are shown in the inset of Fig.~\ref{fig:pyrochlore}(f), and their suppression is an indicator of significant site mixing between the material's $A$ and $B$ sublattices. In the limit where $A$ and $B$ become indistinguishable (\textit{i.e.} perfect site mixing), the cations are effectively sharing a single crystallographic site, and the superlattice peaks disappear entirely. The resulting structure is known as a defect fluorite, with a characteristic $A$O$_{2-x}$ stoichiometry, and described in space group $Fm\overline{3}m$ (no. 225), a supergroup of $Fd\overline{3}m$. The cation site in a defect fluorite has cubic oxygen coordination where one-eighth of the oxygen positions are randomly vacant, while in the pyrochlore structure, these vacancies are site-ordered.

To properly characterize the degree of site mixing in Y\textsubscript{2}(Ti,Zr,Hf,Sn)\textsubscript{2}O\textsubscript{7} its synchrotron powder diffraction pattern was refined in the $Fd\overline{3}m$ space group, with site mixing allowed between the $16c$ and $16d$ Wyckoff positions, \textit{i.e.} the cations were allowed to drift between the lattice sites. To reduce the number of free parameters in the refinement, the high entropy site was modeled by using Pd, the element whose atomic number exactly matches the average of the cation mixture, and hence is a good proxy for the average atomic form factor. The total occupation of each cation crystallographic site and total Y:Pd ratio were also introduced as constraints in the Rietveld refinement. Thus, the effective model corresponds to (Y\textsubscript{2-x}Pd\textsubscript{x})(Pd\textsubscript{2-x}Y\textsubscript{x})O\textsubscript{7} in the pyrochlore structure's typical notation. Using this notation, $x\approx0$ reflects a material whose actual structure is akin to a pyrochlore with minimal site mixing, while $x\approx1$ implies complete site mixing, such that the material is best thought of as a defect fluorite structure. 

This model achieved excellent agreement with the experimental diffraction pattern as shown in Fig.~\ref{fig:pyrochlore}(f) where the refined parameters are presented in Table~\ref{tab:Y2Pd2O7refinement}. The calculated pattern properly accounts for the observed suppression of the (111) and (311) superlattice reflections with good reliability factors (\textcolor{black}{$R_{wp} = 12.01\%$}). The final refined value of $x$ was 0.88, corresponding to a 56:44 mixture of cations on each crystallographic site, extremely close to the fully random 50:50 limit of a defect fluorite structure. Accordingly, a refinement conducted using the latter structure as a model presents nearly identical reliability factors (\textcolor{black}{$R_{wp} = 12.11\%$}), but entirely misses the superlattice peaks. These superlattice peaks represent a minimal fraction of the diffraction pattern's total area, which is why they do not significantly affect the refinement's reliability factors. With these observations in mind, one can conclude that this material is best thought of as being close to the limit of a defect fluorite, with minimal site selectivity as shown in Fig.~\ref{fig:pyrochlore}(d,e), compared to the original target of a site-ordered pyrochlore structure.

The relationship between the pyrochlore and defect fluorite structures has been extensively studied for various combinations of cations in $A$\textsubscript{2}$B$\textsubscript{2}O\textsubscript{7} family materials \cite{liu2004pyrochlore,cepeda2018cations,oquinn2021defining,Fuentes2018}.  Of the constituent cations, it's interesting to note that three of the four tetravalent cations considered here, Ti, Zr, and Sn, are known to form a site-ordered pyrochlore with Y, while only Hf forms a defect fluorite structure. More generally, cation ordering is typically observed to break down in cases where the $A$ and $B$ cations become too similar in size. For example, in Y$_2B_2$O$_7$ solid solutions in which two cations share the $B$-site, a clear phase boundary between pyrochlore and defect fluorite structures is established based on geometric factors that depend only on the average ionic radii of the constituent elements~\cite{oquinn2021defining}. This same geometric factor, when applied to Y$_2$(Ti,Zr,Hf,Sn)$_2$O$_7$ would predict an ordered pyrochlore phase, demonstrating that at higher levels of configurational disorder, conventional stability thresholds can break down. 

\textcolor{black}{As in the perovskite case, the high entropy substitution's effects also manifest in the thermal displacement parameters of our high entropy defect fluorite compound. However, the effect in this case is much more dramatic, resulting in an order of magnitude difference for the cation sublattices of the high entropy and non-disordered reference compounds, a likely consequence of the additional disorder introduced by the site-mixing between the $A$ and $B$ sublattices in the high entropy case. We additionally observe a dramatic 25$\times$ increase in the thermal parameter for the oxygen O2 $8b$ parameter but not the O1 $48f$ site. This too can be directly attributed by the incipient defect fluorite phase as pyrochlores possess a perfect site ordering of the $8b$ and $8a$ oxygen positions, where the latter is completely vacant. In defect fluorites, both sites are randomly occupied with equal probability. This disordering tendency is captured by the large thermal parameter. %This observation emphasizes the structural consequences of the disappearance of site selectivity upon the introduction of chemical disorder in Y\textsubscript{2}(Ti,Zr,Hf,Sn)\textsubscript{2}O\textsubscript{7}.
}

{
\renewcommand{\belowrulesep}{0pt}
\renewcommand{\aboverulesep}{0pt}

{
\setlength{\tabcolsep}{0pt}

\begin{table}[tbp]

\caption{Refined structural parameters for  \textcolor{black}{Y$_2$Sn$_2$O$_7$ and} Y\textsubscript{2}(Ti,Zr,Hf,Sn)\textsubscript{2}O\textsubscript{7} modeled as Y\textsubscript{2}Pd\textsubscript{2}O\textsubscript{7} in the $Fd\overline{3}m$ space group \textcolor{black}{from synchrotron diffraction data collected at ID-28-1 at $T=10$~K.}}
\centering
%\resizebox{\columnwidth}{!}{%
\renewcommand{\arraystretch}{1.1} % Increases row height by 50%
\begin{tabular}{lcccccc}
\toprule
\rowcolor[HTML]{DADADA} 
\multicolumn{1}{c}{\cellcolor[HTML]{DADADA}} &
  \cellcolor[HTML]{DADADA} &
  \multicolumn{3}{c}{\cellcolor[HTML]{DADADA}\textbf{Coordinates}} &
  \cellcolor[HTML]{DADADA} &
  \cellcolor[HTML]{DADADA} \\
\rowcolor[HTML]{DADADA} 
\multicolumn{1}{c}{\multirow{-2}{*}{\cellcolor[HTML]{DADADA}\textbf{ Atom }}} &
  \multirow{-2}{*}{\cellcolor[HTML]{DADADA}\textbf{ Site }} &
  \textbf{x} &
  \textbf{y} &
  \textbf{z} &
  \multirow{-2}{*}{\cellcolor[HTML]{DADADA}\textbf{Occ. }} &
  \multirow{-2}{*}{\cellcolor[HTML]{DADADA}\textbf{\( \mathbf{U_{iso}}  \)(\AA \(\mathbf{^2}\)) \hspace{1pt} } } \\ \hline
  \multicolumn{7}{c}{Y$_2$Sn$_2$O$_7$~~~$a = 10.3938(1)$~\AA} \\ \hline
Y1 &       $16c$               &   0                          &          0                   &  0                           & 1  & 0.0027(2)    \\
\rowcolor[HTML]{DADADA} 
Sn1 &  $16d$                    &  0.5                           &     0.5                        &  0.5                           & 1  & 0.0018(2)    \\

O1 & $48f$                   & 0.344(1)~                    & 0.125~                    & ~0.125~                    & 1        & 0.007(1)  \\
\rowcolor[HTML]{DADADA} 
O2 & $8b$                  & 0.375                    & 0.375                   & 0.375                    & 1        & 0.008(3) \\ \bottomrule
\multicolumn{7}{c}{~~}\\
\toprule
\rowcolor[HTML]{DADADA} 
\multicolumn{1}{c}{\cellcolor[HTML]{DADADA}} &
  \cellcolor[HTML]{DADADA} &
  \multicolumn{3}{c}{\cellcolor[HTML]{DADADA}\textbf{Coordinates}} &
  \cellcolor[HTML]{DADADA} &
  \cellcolor[HTML]{DADADA} \\
\rowcolor[HTML]{DADADA} 
\multicolumn{1}{c}{\multirow{-2}{*}{\cellcolor[HTML]{DADADA}\textbf{ Atom }}} &
  \multirow{-2}{*}{\cellcolor[HTML]{DADADA}\textbf{ Site }} &
  \textbf{x} &
  \textbf{y} &
  \textbf{z} &
  \multirow{-2}{*}{\cellcolor[HTML]{DADADA}\textbf{Occ. }} &
  \multirow{-2}{*}{\cellcolor[HTML]{DADADA}\textbf{\( \mathbf{U_{iso}}  \)(\AA \(\mathbf{^2}\)) \hspace{1pt} } } \\ \hline
  \multicolumn{7}{c}{Y$_2$(Ti,Zr,Hf,Sn)$_2$O$_7$~~~$a = 10.3530(5)$~\AA} \\ \hline
Y1 &                      &                             &                             &                             & 0.44(3)  & 0.011(2)    \\
Pd1 & \multirow{-2}{*}{$16c$} & \multirow{-2}{*}{0}     & \multirow{-2}{*}{0}     & \multirow{-2}{*}{0}     & 0.56(3)  & 0.011(2)    \\
\rowcolor[HTML]{DADADA} 
Y2 &                      &                             &                             &                             & 0.56(3)  & 0.011(2)    \\
\rowcolor[HTML]{DADADA} 

Pd2 & \multirow{-2}{*}{$16d$} & \multirow{-2}{*}{0.5}     & \multirow{-2}{*}{0.5}     & \multirow{-2}{*}{0.5}     & ~0.44(3)  & 0.011(2)    \\

O1 & $48f$                   & 0.344(1)                    & 0.125                    & 0.125                    & 1        & 0.008(4)  \\
\rowcolor[HTML]{DADADA} 
O2 & $8b$                  & 0.375                    & 0.375                    & 0.375                    & 1        & 0.21(3) \\ \bottomrule
\end{tabular}%

\label{tab:Y2Pd2O7refinement}
\end{table}
}
}

\begin{figure}
\centering
\includegraphics[width=\columnwidth]{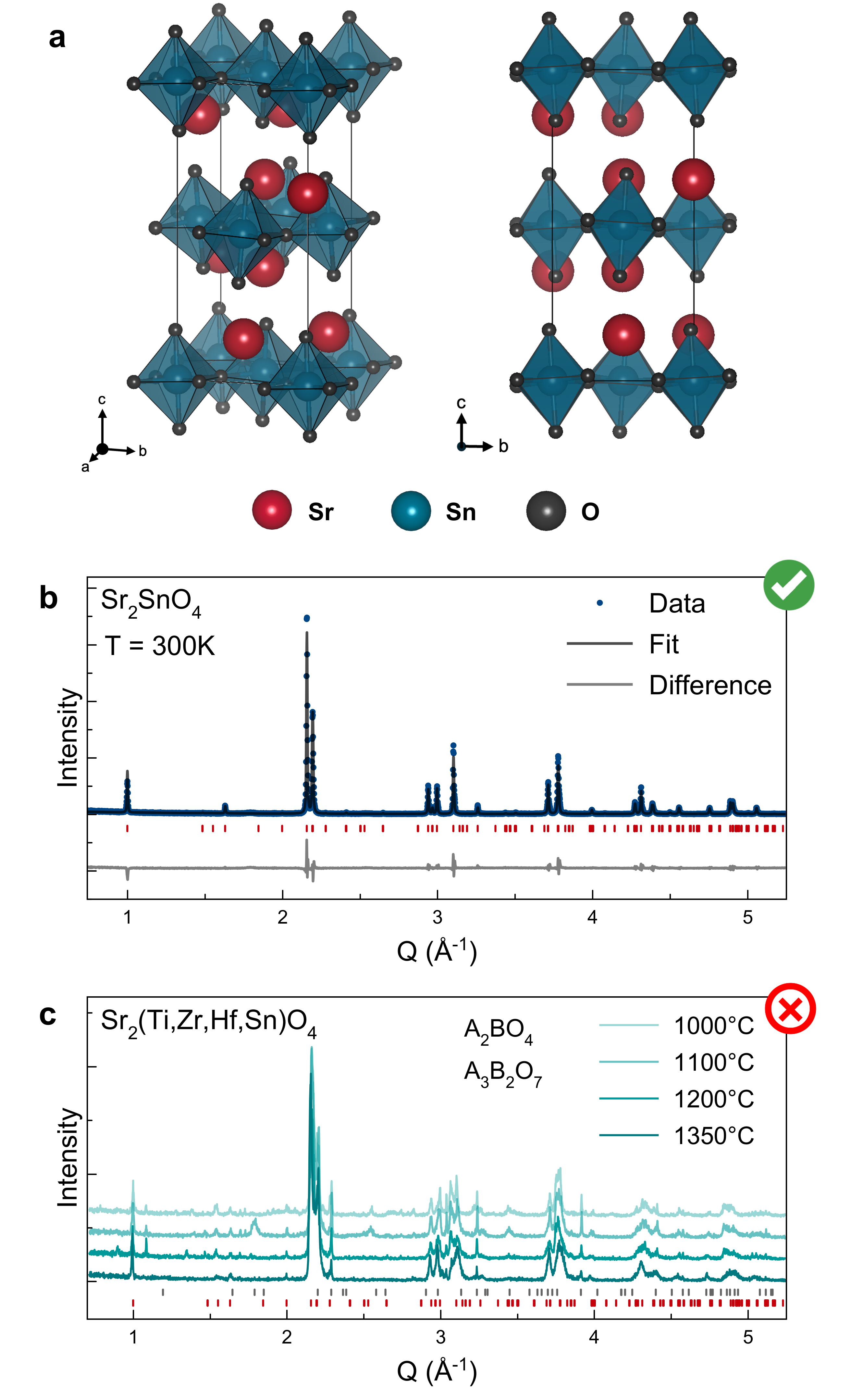}
\caption{\textbf{Instability of a high entropy Ruddlesden-Popper phase.} \textbf{(a)} Crystal structure of Sr$_2$SnO$_4$, which is an orthorhombic Ruddlesden-Popper phase with space group $Pccn$. \textbf{(b)} Rietveld refinement of room temperature laboratory powder XRD data showing the successful solid state synthesis of phase pure Sr$_2$SnO$_4$ ($\chi^2 = 1.71$). \textbf{(c)} XRD patterns for the attempted synthesis of Sr\textsubscript{2}(Ti,Zr,Hf,Sn)O\textsubscript{4} \textcolor{black}{as a function of reaction temperature} showing a large $A_3B_2$O$_7$ impurity that persists to the highest reaction temperatures.}
\label{fig:Ruddlesden}
\end{figure}

Comparing the configurational entropy of a putative pyrochlore (Y$_2$(Ti,Zr,Hf,Sn)$_2$O$_7$) with our observed defect fluorite, (Y$_{0.5}$Ti$_{0.125}$Zr$_{0.125}$Hf$_{0.125}$Sn$_{0.125}$O$_{2-x}$) it is interesting to note that they are identical per disordered site. Even so, the total entropic contribution is substantially larger for the defect fluorite because twice as many sites are affected, leading us to speculate that this phase may be entropy-stabilized. \textcolor{black}{A smaller entropic term arises from the disordering of the oxygen anions across the $8a$ and $8b$ Wyckoff positions, offering additional stability to the defect fluorite phase.} However, the adoption of a defect fluorite phase is not a universal phenomenon: for different cation combinations, high-entropy pyrochlore phases with well-defined $A$ and $B$-sublattices are observed~\cite{teng2020synthesis,jiang2020probing,pitike2022computationally,vayer2023investigation}.

While long-range cation order is precluded by this XRD data, smaller clusters of short-range cation ordering can not be excluded. It is also apparent that the Bragg peaks are substantially broadened in the case of this high-entropy phase, suggesting a high level of lattice strain, likely due to the size mismatch of Y with the smallest tetravalent cations. Likewise, the Bragg peaks present a substantial degree of asymmetry, which becomes particularly apparent at higher $Q$ values. This asymmetry is hard to model with conventional Rietveld refinement techniques, and strongly resembles that observed in Ref.~\cite{Wang_2025_CL_refinement} for a metastable spinel-structured HEO containing a (Mg,Mn,Fe,Cu,Zn) cation mixture. There, it was modeled as a weighted sum of local phases with similar lattice parameters, resulting in asymmetric peak broadening.

\subsection*{Ruddlesden-Popper Structure}

Our third targeted family of materials was the $n=2$ Ruddlesden-Popper, $A_{n+1}B_n$O$_{3n+1}$ phase with the stoichiometry $A_2B$O$_4$. This perovskite-derived structure, which exists in a variety of tetragonal and orthorhombic polytypes, contains a quasi two-dimensional square lattice formed by the $B$-site, as shown in Fig.~\ref{fig:Ruddlesden}(a). For this structure type, our \textcolor{black}{non-disordered reference compound} was Sr\textsubscript{2}SnO\textsubscript{4}, which could be obtained as a phase-pure sample after a single annealing cycle at 1000\textdegree C. This material was found to crystallize in space group $Pccn$ (no. 56) as shown in the Rietveld refinement in Fig.~\ref{fig:Ruddlesden}(b), consistent with previous literature reports~\cite{fu2004neutron}.

In contrast, after annealing the high entropy composition Sr$_2$(Ti,Zr,Hf,Sn)O$_2$ using this same temperature profile, the peaks present in the diffraction pattern could not be attributed to one single phase. Upon refining the XRD pattern, it was evident that the sample was dominated by multiple Ruddlesden-Popper phases including $n=2$ (Sr\textsubscript{2}$B$O\textsubscript{4}) and $n=3$ (Sr\textsubscript{3}$B\textsubscript{2}$O\textsubscript{7}) phases, as well as some evidence of an $n=\infty$ (Sr$B$O\textsubscript{3}) phase. 
In an attempt to isolate a single phase, the sample was subjected to additional heating cycles at incrementally higher temperatures to a maximum of 1350\textdegree C. The temperature evolution of the XRD pattern of Sr$_2$(Ti,Zr,Hf,Sn)O$_4$ is shown in Fig.~\ref{fig:Ruddlesden}(c). The peaks corresponding to both the Sr\textsubscript{2}$B$O\textsubscript{4} and Sr\textsubscript{3}$B\textsubscript{2}$O\textsubscript{7} phases persist after each annealing cycle, and their intensities do not change dramatically. Following the final annealing cycle, it was estimated that a Sr\textsubscript{2}$B$O\textsubscript{4} phase composed approximately 92\% of the sample, whereas the remaining 8\% corresponded to a Sr\textsubscript{3}$B\textsubscript{2}$O\textsubscript{7} phase. \textcolor{black}{We were unable to determine if cation segregation between the observed phases was occurring, due to difficulties in precisely refining the broad and overlapping Bragg peaks.}

This result seems to suggest that diffusion is not the limiting factor in the formation of a high entropy single-phase sample of this material, but rather, thermodynamic stability. Once formed, the stability of the competing Ruddlesden-Popper phase renders the transformation into Sr\textsubscript{2}$B$O\textsubscript{4} phase impossible. \textcolor{black}{Indeed, preiminary attempts to prepare higher order ($n=3$) Ruddlesden-Popper structures with high entropy cation mixtures similarly yields a mixture of phases.} In the present case, given that the Sr$_2B$O$_4$ structure does constitute the majority phase, it is possible that this synthesis could be further optimized towards a single-phase product. An alternative synthesis route, particularly one that enforces higher degrees of cation mixing at the atomic scale (such as combustion or spray pyrolysis), might inhibit the formation of the competing Ruddlesden-Popper phases.

\subsection*{Zirconium Tungstate Structure}

\begin{figure}
\centering
\includegraphics[width=\columnwidth]{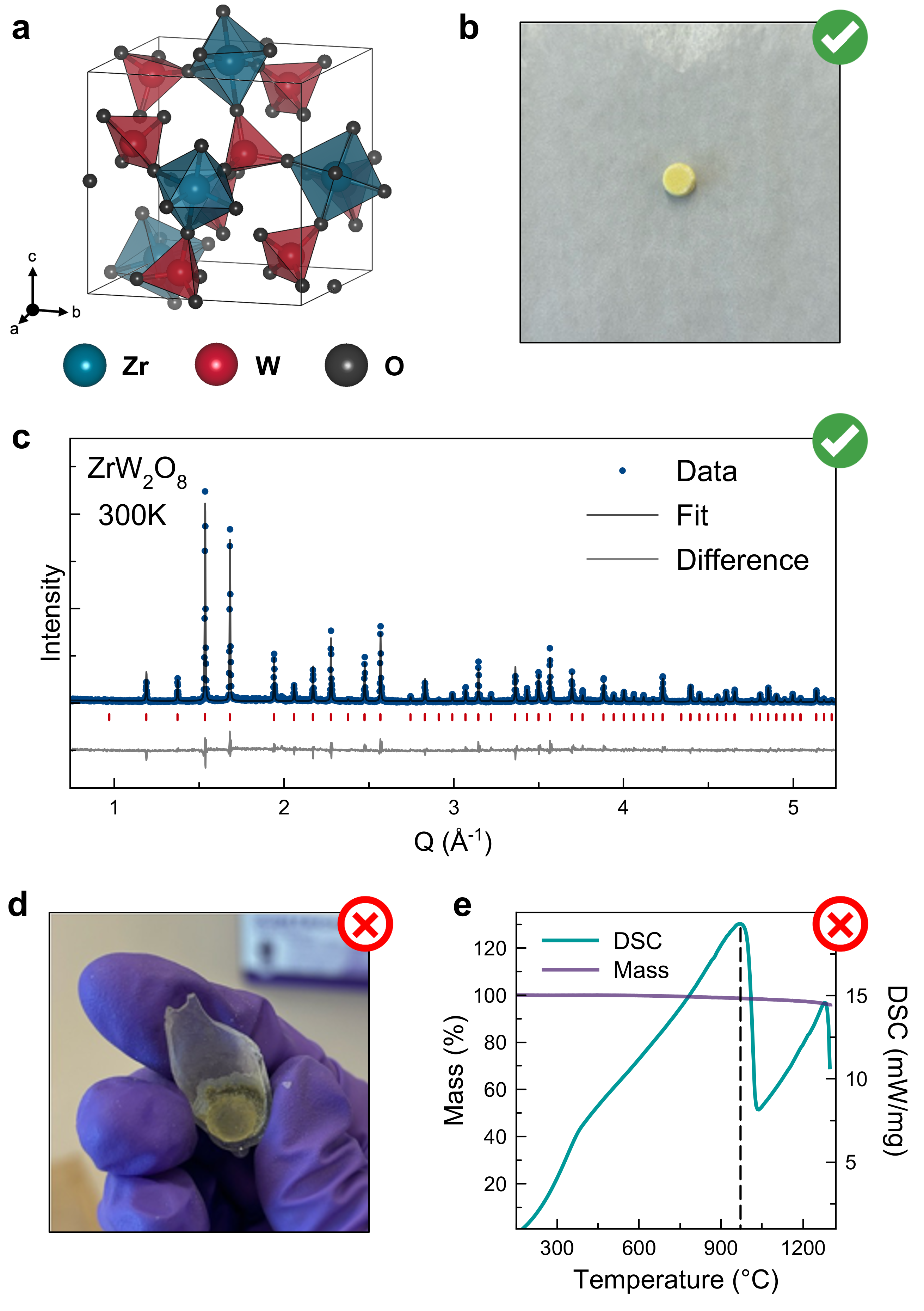}
 \caption{\textbf{Unexpected eutectic point preempts the formation of a high entropy tungstate.} \textbf{(a)} Crystal structure of ZrW\textsubscript{2}O\textsubscript{8}, which is a cubic phase with space group $P2_13$. \textbf{(b)} Photo of resultant ZrW\textsubscript{2}O\textsubscript{8} sample, showing a uniform pellet which has maintained its shape throughout annealing. \textbf{(c)} Rietveld refinement of powder XRD data showing the successful solid state synthesis of phase pure ZrW\textsubscript{2}O\textsubscript{8} ($\chi^2 = 1.60$). \textbf{(d)} Photo of synthesis attempt of (Ti,Zr,Hf,Sn)W\textsubscript{2}O\textsubscript{8} showing the adverse reaction outcome and evidence of melting occurring. \textbf{(e)} Results of DSC-TGA measurement on (Ti,Zr,Hf,Sn)W\textsubscript{2}O\textsubscript{8} precursors, showing a melting phase transition at approximately 970\textdegree C before the target phase can form. }
\label{fig:Tungstate}
\end{figure}

The fourth and final structure investigated here is the zirconium tungstate structure, $A$W\textsubscript{2}O\textsubscript{8}. This structure is only known to form with $A =$~Zr and Hf, both of which are notable negative thermal expansion materials~\cite{OG_Zirc_paper,HfW2O8}. This cubic phase, shown in Fig.~\ref{fig:Tungstate}(a) is composed of corner-sharing WO$_4$ tetrahedra and $A$O$_6$ octahedra, resulting in a highly porous structure with large voids. Our \textcolor{black}{non-disordered reference} phase, ZrW\textsubscript{2}O\textsubscript{8}, formed phase-pure, following a single annealing cycle at 1200\textdegree C, consistent with previous literature reports, and the resulting sample is shown in Fig.~\ref{fig:Tungstate}(b). The powder XRD pattern, shown in Fig.~\ref{fig:Tungstate}(c), was refined in the cubic $P2_13$ (no. 198) space group, showing excellent agreement ($\chi^2=1.60$). 

Initial attempts to obtain a high entropy analogue targeted the synthesis of (Ti,Zr,Hf,Sn)W\textsubscript{2}O\textsubscript{8}, under identical reaction conditions as the \textcolor{black}{non-disordered reference} phase. This synthesis attempt led to evident melting such that the products were fused to the quartz tube reaction vessel, having lost the pellet shape that the initial precursors were pressed into, as shown in Fig.~\ref{fig:Tungstate}(d). A limited amount of product was retrieved and XRD did not reveal any of the diffraction peaks typically associated with ZrW\textsubscript{2}O\textsubscript{8}, suggesting that an analogous phase is not present in the sample. Overall, this XRD pattern was dominated by an amorphous background, and no dominant phases could be clearly identified. As none of the reagents are known to melt at this temperature, an adverse reaction with the container was initially suspected. However, multiple attempts to perform this reaction in different containers, including quartz, alumina, and platinum, gave consistent results of evident melting. 

To investigate the unexpected melting transition, we performed differential scanning calorimetry and thermogravimetric analysis (DSC-TGA) on the  (Ti,Zr,Hf,Sn)W\textsubscript{2}O\textsubscript{8} precursors, which were pre-reacted (Ti,Zr,Hf,Sn)O\textsubscript{2} and WO\textsubscript{3}. The precursors were loaded into an alumina crucible and heated from room temperature to 1300\textdegree C. The result of this DSC-TGA measurement is shown in Fig.~\ref{fig:Tungstate}(e), where a sharp drop in the calorimetry at approximately 970\textdegree C indicates an endothermic reaction that can be assigned as the melting phase transition. Another phase transition is also evident at higher temperature, around 1250\textdegree C, which may be a secondary melting transition.

Informed by the calorimetry result, we next attempted the synthesis of (Ti,Zr,Hf,Sn)W\textsubscript{2}O\textsubscript{8} at temperatures lower than the observed melting point, in 50\textdegree C increments from 800\textdegree C to 950\textdegree C. At each temperature, the sample was annealed for 24 hours, air quenched, and analyzed by XRD. The sample was then reground and pressed into a pellet to be annealed at the next highest temperature increment. It is evident that these annealing temperatures are insufficient in forming the targeted (Ti,Zr,Hf,Sn)W\textsubscript{2}O\textsubscript{8} material, as the precursor oxide structures persisted throughout each of the respective XRD scans, and there was no evidence of a ZrW\textsubscript{2}O\textsubscript{8}-like phase in the XRD patterns. This suggests that the targeted high entropy phase cannot be formed via solid-state reaction.

Given the precursor oxides involved in the synthesis of  (Ti,Zr,Hf,Sn)W\textsubscript{2}O\textsubscript{8} are known to melt at temperatures much higher than 970\textdegree C, it is evident that their mixture is yielding a eutectic point. The origin of this eutectic point was systematically investigated through a process of elimination. We found that the synthesis of a three-component material (Ti,Zr,Hf)W\textsubscript{2}O\textsubscript{8}, as well as the two-component materials (Ti,Zr)W\textsubscript{2}O\textsubscript{8} and  (Ti,Hf)W\textsubscript{2}O\textsubscript{8} all exhibit similar melting reactions when reacted at 1200\textdegree C, while (Zr,Hf)W\textsubscript{2}O\textsubscript{8} and (Zr,Hf,Sn)W\textsubscript{2}O\textsubscript{8} do not melt. We therefore conclude that the reaction between TiO$_2$ and WO$_3$ is primarily responsible for the eutectic melting of the precursors. A binary eutectic between TiO$_2$ and WO$_3$ is known to exist at 1233\textdegree C and 64\% WO$_3$~\cite{chang1967high}. The eutectic uncovered here is significantly deeper, occurring at a temperature that is more than 200\textdegree C lower, and therefore must also pertain to the complex multi-cation environment. Taking these results together, it is evident that forming a high entropy $A$W\textsubscript{2}O\textsubscript{8} phase is challenging, if not impossible. The lack of chemical flexibility in this structure, coupled with its complexity, may contribute to its lack of amenability to hosting a disordered cation sublattice.

%Finally, we note that the compositions that omit Ti ((Zr,Hf)W\textsubscript{2}O\textsubscript{8} and (Zr,Hf,Sn)W\textsubscript{2}O\textsubscript{8}) do not form as phase-pure in the targeted $P2_13$ structure. 
%should be omitted from the reaction in order to mitigate adverse reaction conditions, and ruthenium to be incorporated in its place. Following this, the synthesis of a four component (Zr,Hf,Sn,Ru)W\textsubscript{2}O\textsubscript{8} material was attempted. A phase-pure sample could not be obtained after multiple heating cycles at temperatures up to a maximum of 1250\textdegree C.  While some evidence of a ZrW\textsubscript{2}O\textsubscript{8}-like phase was present in the final XRD pattern, secondary phases were present and did not appear be to be reduced by additional annealing. 
%\textcolor{black}{(not sure if adding specific secondary phases is needed here since the XRD scan isnt included in the figure) The most prevalent secondary phases that could be identified corresponded to a SnO\textsubscript{2}-like rutile phase and a ZrSiO\textsubscript{4} (Zircon)-like phase. [include details about possible reaction with part of quartz vessel leading to a phase like this?] 

\begin{table}[H]
    \centering
          \renewcommand{\arraystretch}{1.2}
    \begin{tabular}{|c|c|}
    \hline
       Compound  & Results \\
       \hline \hline
        BaSnO\textsubscript{3} & Phase-pure  \\
        Ba(Ti,Zr,Hf,Sn)O\textsubscript{3}& Phase-pure  \\ \hline
        Y\textsubscript{2}Sn\textsubscript{2}O\textsubscript{7} & Phase-pure pyrochlore \\
        Y\textsubscript{2}(Ti,Zr,Hf,Sn)\textsubscript{2}O\textsubscript{7} & Phase-pure defect fluorite\\ \hline
        Sr\textsubscript{2}SnO\textsubscript{4} & Phase-pure \\
        Sr\textsubscript{2}(Ti,Zr,Hf,Sn)O\textsubscript{4}& Mixed phase\\ 
        Sr\textsubscript{2}(Ti,Zr,Hf,Sn,Ru)O\textsubscript{4}& Mixed phase\\ \hline

        ZrW\textsubscript{2}O\textsubscript{8} &  Phase-pure
        \\
        
        (Zr,Ti)W\textsubscript{2}O\textsubscript{8} & Melted
        \\(Hf,Ti)W\textsubscript{2}O\textsubscript{8} & Melted
        \\(Zr,Hf,Ti)W\textsubscript{2}O\textsubscript{8} & Melted
        \\(Zr,Hf,Ti,Sn)W\textsubscript{2}O\textsubscript{8} & Melted
        \\(Zr,Hf)W\textsubscript{2}O\textsubscript{8} & Incomplete reaction %\cite{WANG20162837}
        \\(Zr,Hf,Sn)W\textsubscript{2}O\textsubscript{8} & Incomplete reaction
      %  \\%(Zr,Hf,Sn,Ru)W\textsubscript{2}O\textsubscript{8} & Mixed phase
        \\ \hline

    \end{tabular}
    \caption{Summary of syntheses attempted in this study.}
    \label{tab:summary}
\end{table}

\section*{Summary and Outlook}

This work explored the effect of introducing ``high entropy'' –– \textit{i.e.} extreme chemical disorder –– on the synthesizability of complex oxides. Employing a cation combination (Ti, Zr, Hf, Sn) known to stabilize as a single-phase HEO in the absence of other constituents, we explored how configurational entropy, enthalpic effects, and structural constraints interact by attempting high entropy substitution in four structural archetypes. Within these well-characterized materials families, a spectrum of outcomes was observed as summarized in Table~\ref{tab:summary}: (i) the formation of a single-phase in the same structure as the parent compound in $AB$O\textsubscript{3} perovskites; (ii) the adoption of a higher symmetry structure with enhanced configurational entropy resulting in a shift from a site-ordered $A$\textsubscript{2}$B$\textsubscript{2}O\textsubscript{7} pyrochlore to a site-mixed $A$O\textsubscript{2-x} defect fluorite; (iii) the formation of a mixture of phases that are thermodynamically trapped in $AB_2$O$_4$ Ruddlesden-Popper phases; and (iv) the discovery of a multi-cation deep eutectic point such that melting preempts phase formation in the $A$W\textsubscript{2}O\textsubscript{8} tungstate phase. 

This variability in outcomes, despite the shared cation mixture, underscores the complex interdependence of entropic and enthalpic factors that govern the stability of any hypothetical HEO. This stability is influenced by individual cation interactions as well as features dictated by the host structure, such as dimensionality, packing density, local symmetries, and structural rigidity. The combination of these factors will shape the final strain that the high entropy substitution places on the host structure. Notably, within our target structures, the ones with lower symmetry and lower packing efficiencies were generally less stable, while high symmetry, high density structures presented higher stability.  %. Whilst a perovskite HEO forms 
Further work is therefore needed to understand whether this trend is broadly applicable. Crucially, the introduction of high entropy complicates the application of previously-derived empirical proxies used to describe phase-stability, such as tolerance factors, and requires a rethinking of how new HEOs are designed. %Introducing these additional considerations is critical to the targeted design of HEOs –– one of the typical objectives of the field –– when it pertains to basic as well as applied and engineering materials science. 

As we aim to unravel the underlying factors that govern the formation of high entropy materials, several open questions come into view, which are ripe for exploration in future studies. \textcolor{black}{The first is the role of dimensionality in stabilizing an HEO. The phases successfully formed here are all fully three-dimensional. A high entropy mixture on a two-dimensional sublattice will necessarily have inequivalent in-plane and out-of-plane lattice strains, which may give less flexibility for the local distortions that HEOs are known to possess~\cite{jiang2020probing,entropy_mix,jiang2025exploring}. Similar thinking can be applied to low-density structures, regardless of dimensionality: voids in the structure may allow enthalpically-pressured cations to move within the crystallographic arrangement, relieving strain but potentially destabilizing the phase altogether. Finally, we can consider the overall level of structural complexity, which can be approximately quantified by the number of unique cation sublattices. As shown here, each additional new constituent brings with it the risk of an adverse pairwise interaction, which may preempt phase formation. We therefore speculate that reduced dimensionality, reduced packing density, and increased structural complexity may all be unfavorable for the formation of an HEO.} 

Overall, our understanding of HEO stability remains in its infancy. The sheer combinatorics of HEO design render blind searches through their configurational space ineffective and inefficient. Therefore, efforts to understand the factors governing successful phase formation can inform the new set of design principles and heuristics that are needed to explore this vast phase space.  

%the importance of testing synthetic routes that bypass diffusional and enthalpic barriers to phase formation, such as hydrothermal, hydroflux, and high pressure syntheses cannot be overstated. Their exploration provides invaluable information to establishing bounds on the stability of high entropy systems, while highlighting potentially overlooked connections between structural and chemical features, as a wider slice of phase space is varied. 

%

\section*{Methods}

\noindent \textbf{Synthesis of HEO powders:} Polycrystalline powder samples of each material were obtained through solid-state synthesis. \textcolor{black}{The respective precursors were weighed in their stoichiometric ratios, ground together in an agate mortar and pestle, and pressed into a pellet with the use of a hydraulic press. The carbonate precursors SrCO\textsubscript{3} and BaCO\textsubscript{3} were used in the synthesis of perovskite and Ruddlesden-Popper structures, whereas all other precursors used were in their binary oxide form.}
Each reaction occurred in air, with reactants held in either a quartz tube or alumina crucible reaction vessel, chosen to suit specific reaction conditions.
The reactants were placed into a \textit{Thermo Scientific - Lindberg Blue M} high temperature furnace, and reacted at an appropriate temperature between 1100\textdegree C and 1400\textdegree C for at least 24 hours. \textcolor{black}{Each sample was heated at a rate of 100\textdegree C per hour, and was air-quenched once the reaction time had elapsed.}
In cases where the samples were not phase-pure after one reaction cycle, the pellet was reground and re-heated, typically at a slightly higher temperature, chosen based on the constituents' phase diagrams. The final synthesis temperatures were  1000\textdegree C for Sr\textsubscript{2}SnO\textsubscript{4}, 1200\textdegree C for ZrW\textsubscript{2}O\textsubscript{8}, 1300\textdegree C for BaSnO\textsubscript{3} and Ba(Ti,Zr,Hf,Sn)O\textsubscript{3}, and 1350\textdegree C for Sr\textsubscript{2}(Ti,Zr,Hf,Sn)O\textsubscript{4}, Y\textsubscript{2}Sn\textsubscript{2}O\textsubscript{7}, and Y\textsubscript{2}(Ti,Zr,Hf,Sn)\textsubscript{2}O\textsubscript{7}. \\

\noindent \textbf{Laboratory XRD:} Powder XRD measurements of each sample were taken using a Bruker D8 ADVANCE diffractometer equipped with a Cu x-ray source and Johansson monochromator emitting a monochromatic beam wavelength of $\lambda_{K\alpha_1}=1.5046$~\AA. Rietveld refinements for each scan were carried out using TOPAS~\cite{coelho2018topas}. Refined quantities include the background, sample displacement, lattice constants, atomic positions, modified peak shape, strain broadening, and thermal parameters.\\

\noindent \textbf{Synchrotron XRD:} For samples with $AB$O\textsubscript{3} and $A$\textsubscript{2}$B$\textsubscript{2}O\textsubscript{7} stoichiometries, medium-resolution synchrotron powder diffraction experiments were carried out at beamline 28-ID-1 (PDF) of the National Synchrotron Light Source II (Brookhaven National Laboratory, Upton, NY). Samples were loaded into polyimide capillary tubes. An incident wavelength of 0.1665 \AA was used, and diffraction data were collected using an amorphous silicon area detector (PerkinElmer, Waltham, MA), placed 216.7 mm behind the samples. The sample temperature was maintained at 10 K using a closed-cycle cryostat, and calibration of the experimental setup was done using a nickel powder beamline standard.\\

\noindent \textbf{DSC-TGA:} Differential scanning calorimetry (DSC) and thermogravimetric analysis (TGA) measurements were carried out using a Netzsch Simultaneous Thermal Analysis (STA) 449 F5 Jupiter instrument. Measurements were taken between room temperature and 1300\textdegree C. Data were collected with a ramping rate of 20~K/min, under flowing argon.

\section*{Data availability}
The data that support the findings of this study are available from the corresponding author upon reasonable request.

\begin{acknowledgments}
The authors thank Dr. Milinda Abeykoon, Dr. Solveig S. Aamlid, and Allison Pavlik for help carrying out the synchrotron diffraction experiments and Dr. Austin Ferrenti for pointing out the relationship between density and phase stability. This research used beamline 28-ID-1 of the National Synchrotron Light Source II, a U.S. Department of Energy (DOE) Office of Science User Facility operated for the DOE Office of Science by Brookhaven National Laboratory under Contract No. DE-SC0012704. This work was supported by the Natural Sciences and Engineering Research Council of Canada (NSERC), the Canadian Institute for Advanced Research (CIFAR), the Sloan Research Fellowships program, and the Killam Accelerator Research Fellowship. KK was supported by an NSERC Undergraduate Student Research Award. This research was undertaken thanks in part to funding from the Canada First Research Excellence Fund, Quantum Materials and Future Technologies Program.
\end{acknowledgments}

%\section*{Appendix}

%\subsection*{Synthesis Results}

%\subsection*{Experimental data for refined structures}

%\newpage
\bibliography{refs}

%\begin{figure*}
%  \centering
 % \includegraphics[width=3.25in]{TOC_Image_2.png}
 % \par For Table of Contents Only
  %\label{fgr:TOC}
  %\end{figure*}

\end{document}